\documentclass[aps,pra,reprint]{revtex4-2}

\usepackage{custom_aps}
\PassOptionsToPackage{full}{textcomp}	% required for newtx
\usepackage{newtxtext,newtxmath}		% fonts
\crefname{section}{Sec.}{Secs.}			% correct references for sections
\Crefname{section}{Section}{Sections}	% correct references for sections
\crefname{figure}{Fig.}{Figs.}			% correct references for figures
\Crefname{figure}{Figure}{Figures}		% correct references for figures
\crefrangelabelformat{subfigure}{#3#1#4--#5#2#6}	% required reference format

\usepackage[caption=false]{subfig}		% subcaption incompatible with RevTeX
\newcommand{\phantomsubfloat}[1]{
	\captionsetup[subfigure]{labelformat=empty}	% apply caption setup only temporarily
	\subfloat[][]{#1}
}

\pdfoutput=1

\definecolor{linkcolour}{rgb}{0.02,0.12,0.3}
\hypersetup{colorlinks=true,citecolor=linkcolour,linkcolor=linkcolour,urlcolor=linkcolour}

\newcommand{\ncrit}{n_\text{2D}^\text{crit}}
\newcommand{\ntwoD}{n_\text{2D}}

\begin{document}
	
	\title{How to realise a homogeneous dipolar Bose gas in the roton regime}
	
	\begin{abstract}
		Homogeneous quantum gases open up new possibilities for studying many-body phenomena and have now been realised for a variety of systems. For gases with short-range interactions the way to make the cloud homogeneous is, predictably, to trap it in an ideal (homogeneous) box potential. We show that creating a close to homogeneous dipolar gas in the roton regime, when long-range interactions are important, actually requires trapping particles in soft-walled (inhomogeneous) box-like potentials. In particular, we numerically explore a dipolar gas confined in a pancake trap which is harmonic along the polarisation axis and a cylindrically symmetric power-law potential $r^p$ radially. We find that intermediate $p$'s maximise the proportion of the sample that can be brought close to the critical density required to reach the roton regime, whereas higher $p$'s trigger density oscillations near the wall even when the bulk of the system is not in the roton regime. We characterise how the optimum density distribution depends on the shape of the trapping potential and find it is controlled by the trap wall steepness.
	\end{abstract}
	
	\author{Péter Juhász}
	\thanks{P.\ J.\ and M.\ K.\ contributed equally to this work.}
	\author{Milan Krstaji\'{c}}
	\thanks{P.\ J.\ and M.\ K.\ contributed equally to this work.}
	\author{David Strachan}
	\author{Edward Gandar}
	\author{Robert P. Smith}
	\email{robert.smith@physics.ox.ac.uk}
	
	\affiliation{Clarendon Laboratory, University of Oxford, Parks Road, Oxford, OX1 3PU, United Kingdom}
	
	\date{30\textsuperscript{th} May, 2022}
	
	\maketitle
	
	The behaviour of many-body quantum systems is governed by the interplay of the potential confining the particles and the interactions between them; ultracold gases allow for the fine control of both of these aspects. While in most ultracold-atom experiments interparticle interactions are short-ranged and isotropic, the realisation of ultracold dipolar gases, using highly magnetic atoms~\cite{Griesmaier:2005, Lu:2011, Aikawa:2012, Chomaz:2022}, molecules~\cite{Bohn:2017} and Rydberg atoms~\cite{Saffman:2010}, has introduced anisotropic, long-range dipole--dipole interactions, opening up many new avenues for research. In the case of degenerate Bose gases, the presence of dipole--dipole interactions has, for example, led to the study of roton physics~\cite{Santos:2003, Chomaz:2018} and the related discovery of a supersolid phase~\cite{Bottcher:2019, Tanzi:2019, Chomaz:2019}.
	
	The term `roton' was first coined in the context of liquid helium~\cite{Landau:1947}, where it describes excitations observed around a minimum in the excitation spectrum at nonzero momentum. Ultracold dipolar gases tightly confined along the polarisation direction of the dipoles and held more loosely in (at least one of) the other two directions display a similar roton dispersion relation. In this case, the origin of the roton feature is the interplay of the anisotropic, long-range interactions and the tight confinement. As the strength of the interactions is increased, the roton minimum forms, deepens and then reaches zero energy, causing the roton instability. In certain cases, this leads to the formation of quantum droplets~\cite{Kadau:2016, Ferrier-Barbut:2016, Chomaz:2016} and, very close to the instability, a supersolid phase~\cite{Bottcher:2019, Tanzi:2019, Chomaz:2019, Bottcher:2021}.
	
	In the experiments so far, the dipolar gases were confined in anisotropic, harmonic potentials; theoretically, most attention has focused on such fully harmonically trapped gases~\cite{Ronen:2007, Wilson:2008, Blakie:2020, Roccuzzo:2020, Zhang:2021, Poli:2021, Gallemi:2020, Natale:2019, Hertkorn:2019, Tanzi:2019b, Guo:2019, Hertkorn:2021, Hertkorn:2021b, Ilzhofer:2021, Norcia:2021, Tengstrand:2021, Bland:2021} and on homogeneous condensates~\cite{Baillie:2015, Ancilotto:2019, Zhang:2019, Blakie:2020b, Ancilotto:2021, Turmanov:2021, Pal:2022} which are harmonically confined along the polarisation direction but are unconfined in at least one of the other two (in-plane) directions. The natural way to create homogeneous conditions experimentally is to make the in-plane confinement box-like. Box traps had much success in systems with purely contact interactions~\cite{Navon:2021}, as making a condensate homogeneous almost invariably makes the interpretation of experiments easier and the comparisons with theory more direct. Such traps are yet to be used for experimentally studying many-body phenomena in dipolar quantum gases, but theoretical studies involving ideal box traps have revealed non-trivial effects such as the accumulation of density near the box walls~\cite{Lu:2010} and novel supersolid crystal structures~\cite{Roccuzzo:2022}.
	
	\begin{figure}[b]
		\setlength{\tabcolsep}{0pt}
		\centering
		\begin{tabular}[c]{rl}
			\parbox[c]{0.28\columnwidth}{\raggedleft\includegraphics[width=0.28\columnwidth]{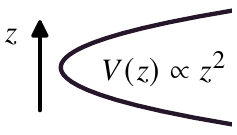}}&
			\parbox[c]{0.5\columnwidth}{\raggedright\includegraphics[width=0.5\columnwidth]{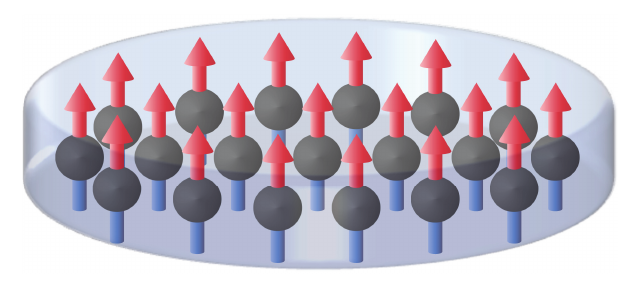}}\\
			&
			\parbox[c]{0.5\columnwidth}{\raggedright\includegraphics[width=0.5\columnwidth]{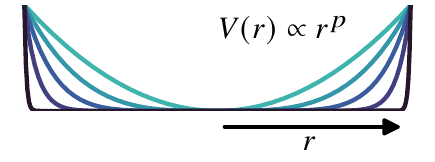}}\\
		\end{tabular}
		\caption{Trap geometry. We consider a gas of dipoles aligned along $z$ which interact via dipole--dipole and contact interactions, and are confined in a `pancake' trap by a harmonic potential along the axis ($z^2$) and a cylindrically symmetric power-law potential in the radial direction ($r^p$).}
		\label{fig:setup}
	\end{figure}
	
	In this Letter, we numerically explore the homogeneity of a dipolar gas, tuned close to the roton instability, in a flattened, cylindrically symmetric (`pancake') potential, with tight harmonic confinement along $z$ (the direction of polarisation of the dipoles) and a power-law potential $r^p$ in the perpendicular plane (see \cref{fig:setup}). This choice is motivated by the fact that a general power-law potential smoothly interpolates between a harmonic potential ($p=2$) and an ideal box potential ($p \to \infty$), and that experimentally relevant box-like traps are typically characterised as power-law potentials~\cite{Gaunt:2013}. Additionally, Laguerre--Gaussian beams, often used to create optical box traps, can be used to controllably realise power-law potentials. We should emphasise that we are not looking for the most homogeneous system with any dipolar interaction strength, but we are exploring how closely one can replicate an infinite homogeneous system at the critical boundary for the roton instability. While one cannot realise infinitely large systems experimentally, recreating the same conditions in an extended but finite region should still reproduce infinite homogeneous system phenomena. We find that achieving the most homogeneous conditions within the roton regime requires an intermediate $p$ which depends on the trap aspect ratio---this is in sharp contrast to systems with only contact interactions, where a higher $p$ always leads to more homogeneous condensates~\cite{Navon:2021}. We show that the optimum $p$ for a given aspect ratio is determined by the box walls being soft enough so as not to trigger a roton-like instability at the edge significantly before it occurs in the bulk. We also examine how the optimum $p$ depends on the aspect ratio and how homogeneous-system-like a sample could be produced within realistic experimental limitations.
	
	We consider a bosonic gas of $N$ atoms, each with mass $m$ and magnetic dipole moment $\mu_m$, confined in a pancake trap (see \cref{fig:setup}), and we work in dimensionless units where times are expressed in units of the inverse $z$-axis oscillator frequency $1/\omega_z$, energies in units of $\hbar\omega_z$ and lengths in units of the harmonic oscillator length $\ell_z=\sqrt{\hbar/(m\omega_z)}$. The trapping potential we are considering is given by
	\begin{equation}
		V_{\text{trap}}(r,z)= \frac{1}{2} \left( \left( \frac{r}{\gamma}\right)^p + z^2 \right),
	\end{equation}
	where $\gamma$ characterises the aspect ratio of the trap.
	
	At zero temperature, our dipolar gas is expected to form a Bose--Einstein condensate (BEC) described by the macroscopic wave function $\Psi(\vb{r},t)=\sqrt{N}\psi(\vb{r},t)$, where $\psi(\vb{r},t)$ is the normalised single-particle wave function that obeys the (dimensionless) Gross--Pitaevskii equation (GPE):
	\begin{multline}
		i \frac{\partial \psi(\vb{r},t)}{\partial t}= \bigg(-\frac{1}{2}\nabla^2 + V_{\text{trap}}(\vb{r}) + g_s \abs{\psi(\vb{r},t)}^2 \\
		+D \Phi_{\text{dd}}(\vb{r},t)\bigg) \psi(\vb{r},t).
		\label{eq:dimGPE}
	\end{multline}
	Here, the strength of contact interactions is characterised by the parameter $g_s=4\pi a_s N/\ell_z$, where $a_s$ is the s-wave scattering length, and the strength of dipolar interactions is characterised by the parameter $D=3 a_{\text{dd}}N/\ell_z$, where $a_{\text{dd}}=m\mu_0\mu_m^2/(12\pi\hbar^2)$ is the dipolar length and $\mu_0$ is the permeability of free space. The form of the mean-field dipolar interaction potential is given by
	\begin{equation}
		\Phi_{\text{dd}}(\vb{r},t) = \int \frac{ 1-3 \cos^2 \theta}{\abs{\vb{r}-\vb{r'}}^3} \abs{\psi(\vb{r'},t)}^2 \dd[3]{\vb{r'}},
	\end{equation}
	where $\theta$ is the angle between $\vb{z}$ and $\vb{r}-\vb{r'}$. The relative strength of these interactions (compared to the contact interactions) is given by the ratio $\varepsilon_\text{dd}=a_\text{dd}/a_s$. Note that here we neglect quantum fluctuations, as their contribution is negligible up to the typical densities required for mean-field collapse; they only become significant at higher densities when they can arrest the collapse, leading to quantum droplets or supersolidity~\cite{Ferrier-Barbut:2016}.
	
	In the absence of an in-plane potential (i.e.\ $p \to \infty$ and $\gamma \to \infty$), a dipolar gas is predicted to develop a roton-like excitation spectrum, with a roton minimum for excitations of wavelength $\lambda_\text{rot} \approx 2\pi$~\cite{Santos:2003}, which deepens with increasing dipolar interaction strength and reaches zero energy at the roton instability. The instability occurs when the single-particle areal density $\ntwoD(r)=\int_{-\infty}^{\infty} \abs{\psi(\vb{r})}^2\dd{z}$ reaches a critical value~\cite{Santos:2003, Baillie:2015} given by
	\begin{equation}
		\ncrit(\varepsilon_\text{dd})=\frac{3 \nu_\text{crit}(\varepsilon_\text{dd})}{4\pi D(\varepsilon_\text{dd}^{-1}+2)},
		\label{eq:nucrit}
	\end{equation}
	where the value of the dimensionless prefactor $\nu_\text{crit}(\varepsilon_\text{dd})$ is tabulated in Ref.~\cite{Baillie:2015}. Interestingly, up until the gas becomes unstable, the density distribution of the BEC has the same form as a gas with only contact interactions with an effective scattering length $a_\text{eff} = a_s + 2 a_\text{dd} = a_\text{dd} (\varepsilon_\text{dd}^{-1} + 2) $~\cite{Baillie:2015}, or equivalently an effective interaction parameter $g_\text{eff}=4\pi a_\text{eff} N/\ell_z$.
	
	In our simulations, for each trap with given $\{\gamma,p\}$ and for a given $\varepsilon_\text{dd}$, we solve the GPE and find the maximum value of $D$ ($g_s$ is fixed by $g_s=4\pi D/(3\varepsilon_\text{dd})$) for which a stable ground state can be found (see the \hyperref[app:numerics]{Appendix} for further information about our algorithm). As we aim to compare the resulting critical density distributions to the infinite (perfectly homogeneous) flattened system, we evaluate the $r$-dependent areal density $\ntwoD(r)$ and compare it to $\ncrit$, the density a perfectly homogeneous system would have at the roton instability.
	
	\begin{figure}[t]
		\centering
		\vspace{-1\baselineskip}  % Remove extra line inserted by subfloat
		\phantomsubfloat{\label{fig:n2Dexamples_nvsr_2}}
		\phantomsubfloat{\label{fig:n2Dexamples_nvsr_6}}
		\phantomsubfloat{\label{fig:n2Dexamples_nvsr_20}}
		\phantomsubfloat{\label{fig:n2Dexamples_nvsP_2}}
		\phantomsubfloat{\label{fig:n2Dexamples_nvsP_6}}
		\phantomsubfloat{\label{fig:n2Dexamples_nvsP_20}}
		\includegraphics[width=1.0\columnwidth]{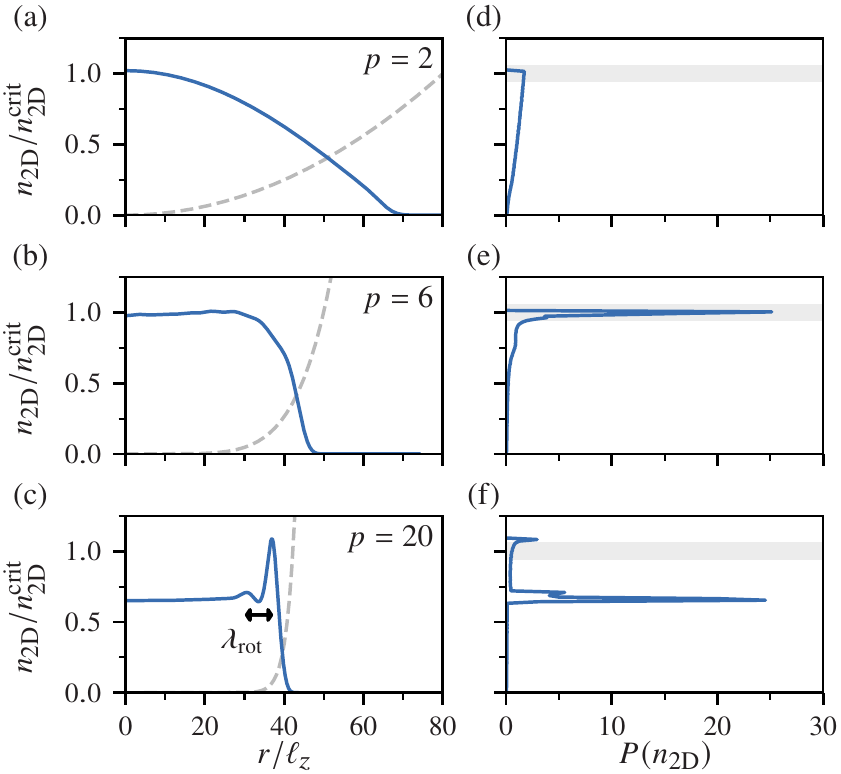}
		\caption{Critical density distributions of a purely dipolar gas. \,(a)--(c)~Areal density distributions $\ntwoD$ (solid blue lines) at the highest interaction strength at which a stable ground state can form in a power-law potential with $\gamma=40$ and $p=2, 6$ and 20, respectively (dashed grey lines). The $\ntwoD$'s are given relative to $\ncrit$, the critical density for the roton instability in an \emph{infinite} flattened system; the trap potential is given relative to the chemical potential and shares the same axis. It can be seen that instability in the trapped system occurs when the maximum $\ntwoD$ is close to $\ncrit$. For $p=20$, this is due to a pronounced density oscillation near the trap wall, whose wavelength is close to $\lambda_\text{rot}$ (see arrow). \,(d)--(f)~Corresponding probability density distributions $P(\ntwoD)$ of the areal density (see text). $P(\ntwoD)$ is plotted on the horizontal axis such that the vertical axis is shared with plots (a)--(c), the grey shading denotes the region within \SI{5}{\percent} of $\ncrit$. Whereas for the $p=6$ trap \SI{63}{\percent} of the atoms are within \SI{5}{\percent} of $\ncrit$, for both high and low $p$ only a small fraction is.}
		\label{fig:n2Dexamples}
	\end{figure}
	
	\Cref{fig:n2Dexamples_nvsr_2,fig:n2Dexamples_nvsr_6,fig:n2Dexamples_nvsr_20} show examples of $\ntwoD(r)/\ncrit$ for a purely dipolar gas ($\varepsilon_\text{dd}\to\infty$) with three different $p$'s for $\gamma=40$. In all cases, the gas becomes unstable when $\ntwoD(r)$ reaches $\ncrit$ (or just above) somewhere in the trap, suggesting the local onset of the homogeneous roton instability (in a local density approximation picture). While for $p=2$ and $p=20$ the critical density is only reached at the trap centre and the trap edge respectively, for $p=6$ it is reached across most of the gas simultaneously. To further highlight this, in \cref{fig:n2Dexamples_nvsP_2,fig:n2Dexamples_nvsP_6,fig:n2Dexamples_nvsP_20} we plot the corresponding probability density distributions $P(\ntwoD)$, where $P(\ntwoD)\dd{\ntwoD/\ncrit}$ gives the probability of finding a particle at a density between $\ntwoD$ and $\ntwoD+\dd{\ntwoD}$. For a perfectly homogeneous system $P(\ntwoD)$ would be a delta function. For $p=2$ we see that $P(\ntwoD)$ varies smoothly and only a small fraction of the particles are near $\ncrit$. The distribution for $p=20$ is very different, with a large peak corresponding to the bulk of the system at $\ntwoD/\ncrit \approx 0.6$, but still with only a small fraction near $\ncrit$. For $p=6$, the peak corresponding to the bulk of the system sits at $\ncrit$ and so the majority of the system approaches the roton instability simultaneously.
	
	\begin{figure}
		\centering
		\vspace{-1\baselineskip}  % Remove extra line inserted by subfloat
		\phantomsubfloat{\label{fig:Hanalysis_Hvsp}}
		\phantomsubfloat{\label{fig:Hanalysis_HvsS}}
		\phantomsubfloat{\label{fig:Hanalysis_pstarvsgamma}}
		\phantomsubfloat{\label{fig:Hanalysis_Hmaxvsgamma}}
		\includegraphics[width=1.0\columnwidth]{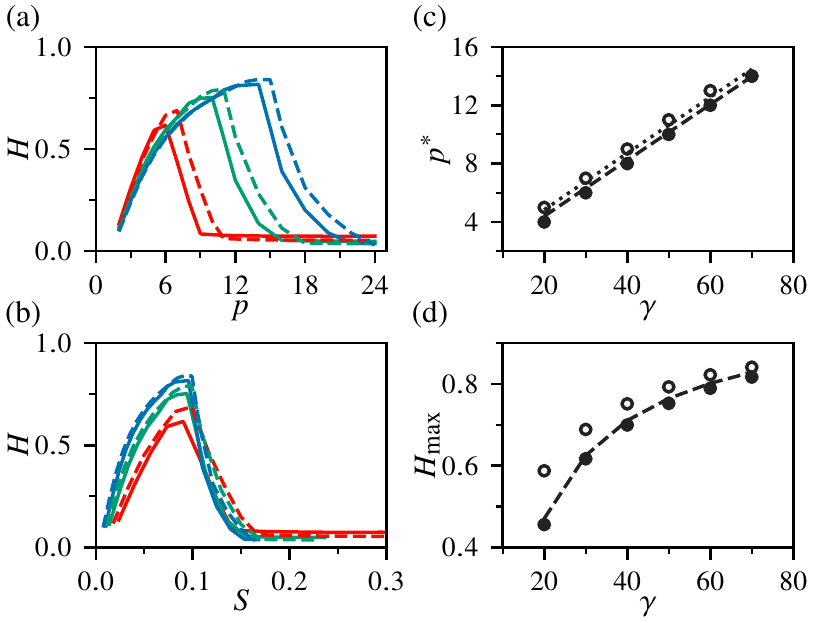}
		\caption{Optimum power-law. \,(a)~The `homogeneity' $H$ (see text) as a function of the power-law exponent $p$ for trapping potential aspect ratio $\gamma=30$ (red, lower lines), 50 (green, middle lines) and 70 (blue, upper lines). The curves are plotted for a purely dipolar gas (solid lines) and for a gas with $\varepsilon_\text{dd}=3$ (dashed lines). The optimum $p(=p^*)$ increases with $\gamma$. \,(b)~The same data plotted against the trap wall steepness $S$ (see text). The optimum $H$ occurs at the same $S \approx 0.1(=S^*)$ for all $\gamma$. \,(c)~The optimum power-law $p^*$ (dots) and the prediction $p^*_\text{pred}$ using $S^*=0.1$ (lines) as a function of $\gamma$ for a purely dipolar gas (filled dots, dashed line) and for one with $\varepsilon_\text{dd}=3$ (empty dots, dotted line). \,(d)~The maximum $H(= H_\text{max})$, achievable for a given $\gamma$ (filled dots for a purely dipolar gas, empty dots for $\varepsilon_\text{dd}=3$). The dashed line provides a simple interpretation of how $H_\text{max}$ depends on $\gamma$ (see text).}
		\label{fig:Hanalysis}
	\end{figure}
	
	We note that the increase of density seen near the trap walls in high-$p$ traps (cf.\ \cref{fig:n2Dexamples_nvsr_20}) is a consequence of the repulsive (and long-range) nature of the interaction between side-by-side dipoles and can be understood in the Thomas--Fermi approximation, in which $V_\text{ext}(\vb{r}) + g_s\abs{\psi(\vb{r})}^2 + D \Phi_{\text{dd}}(\vb{r})$ must be constant and equal to the chemical potential within the cloud. In a sharp-walled trap, the contribution of the external potential is negligible. However, due to the long-range nature of the dipolar interactions, if the gas had a homogeneous density distribution, the dipolar term would be significantly reduced near the wall, so the density needs to increase to compensate. For less steep traps, the increasing $V_\text{ext}$ compensates the decay of $\Phi_{\text{dd}}$ and so no density accumulation occurs near the edge of the trap.
	
	To better quantify the power-law best suited for studying the physics of a homogeneous system in the roton regime, we define a `homogeneity' parameter $H$ as the fraction of particles that experience an $\ntwoD$ within \SI{5}{\percent} of $\ncrit$. We note this parameter quantifies how close the system is to a perfectly homogeneous system \emph{at the roton instability}, and not (only) how uniform the density is across the sample. In \cref{fig:Hanalysis_Hvsp}, we plot $H$ against the exponent $p$ for aspect ratios $\gamma=30, 50$ and 70 for a purely dipolar gas ($\varepsilon_\text{dd} \to \infty$, solid lines). For all three aspect ratios, $H$ gradually increases with $p$ up to some optimum $p^*$ before dropping sharply with higher $p$ as the peak in $P(\ntwoD)$ moves below $0.95\,\ncrit$. We see that $p^*$ increases with $\gamma$; it is determined by the $p$ at which significant density starts accumulating near the edge of the trap (cf.\ \cref{fig:n2Dexamples_nvsr_20}). We have also checked that this behaviour is not specific to purely dipolar gases but also applies in the presence of (weak) contact interactions. We show the curves for $\varepsilon_\text{dd}=3$ (dashed lines), and see that repulsive contact interactions increase both $p^*$ and the maximum $H$ slightly (attractive contact interactions have the opposite effect).
	
	One would expect density oscillations near the wall to somehow be controlled by the trap wall steepness, which not only depends on $p$ but also on $\gamma$. We define the steepness as the gradient of the trap potential (relative to the chemical potential $\mu$) at half the chemical potential:
	\begin{equation}
		S= \dv{(V_\text{trap}/\mu)}{r} \Bigr\rvert_{V_\text{trap}(r,0)=\mu/2} = \frac{p}{2\gamma} \mu^{-1/p}.
		\label{eq:S}
	\end{equation}
	In \cref{fig:Hanalysis_HvsS}, we plot $H$ for the same aspect ratios as in \cref{fig:Hanalysis_Hvsp} but now against $S$. Plotting this way reveals that the maximum $H$ occurs at the same $S = S^* \approx 0.1$ for all three aspect ratios for a dipolar condensate with or without contact interactions. Given $\{p,\gamma,\mu\}$ uniquely defines $S$, we can invert \cref{eq:S} and use $S^*$ to predict
	\begin{equation}
		p^*_\text{pred}=\frac{\ln{\mu}}{W_0\left(\frac{\ln{\mu}}{2 S^* \gamma}\right)},
		\label{eq:p_star}
	\end{equation}
	where $W_0(x)$ is the Lambert $W$ function. To check this, in \cref{fig:Hanalysis_pstarvsgamma} we plot $p^*$ and $p^*_\text{pred}$ for a range of $\gamma$ and see there is very good agreement. This behaviour is in contrast to a gas with only contact interactions ($\varepsilon_\text{dd}=0$), where homogeneity would monotonically increase with $S$, but would saturate when $\delta=1/S$ (the trap `wall thickness') reaches the healing length $\xi=1/\sqrt{2\mu}$ (for our parameters $\xi \approx 0.5 \ll \lambda_\text{rot}$). In our case, we reach the optimum $H$ at $\delta \approx 10 \gg \xi$, which is close to the roton wavelength $\lambda_\text{rot} \approx 2\pi$.
	
	As shown in \cref{fig:Hanalysis_Hmaxvsgamma}, as $\gamma$ increases, the maximum $H$ ($H_\text{max}$, achieved at the also growing $p^*$) increases towards 1, suggesting that the homogeneous limit can still in principle be approached if $\gamma$ and $p$ are increased together in a suitable way. The trend can be understood via a simple model (dashed line). If we assume the cloud consists of a homogeneous centre with radius $\gamma-\lambda_\text{rot}$ and an inhomogeneous boundary with a width $\lambda_\text{rot}$, we can estimate $H_\text{max} \approx (1-\lambda_\text{rot}/\gamma)^2$.
	
	Finally, we consider the implications for experimentally realising a close-to-homogeneous dipolar gas in the roton regime. Unlike for gases with solely repulsive contact interactions, the need for relatively soft walls means that the optics for creating an appropriate trap is unlikely to be a significant constraint. Instead, the limiting factor is likely to be the number of atoms required to fill a high-$\gamma$ trap. For an approximately uniform gas in the roton regime $\ntwoD \approx 1/(\pi \gamma^2) \approx \ncrit$, which using \cref{eq:nucrit} gives $\gamma^2=4N a_\text{eff}/(\nu_\text{crit}(\varepsilon_\text{dd}) \ell_z)$. This shows that filling a large-$\gamma$ trap requires $\ell_z$ to be small; however, $\ell_z$ needs to be kept large enough to avoid high (3D) number densities which result in excessive three-body losses. The (dimensionful) peak density can be obtained via the chemical potential and is given by $n_\text{3D}^\text{max} \approx \mu_h(\varepsilon_\text{dd})/g_\text{eff} \times N/\ell_z^3= \mu_h(\varepsilon_\text{dd})/(4 \pi a_\text{eff} \ell_z^2)$, where $\mu_h(\varepsilon_\text{dd})$ is the (dimensionless) chemical potential tabulated in Ref.~\cite{Baillie:2015}. Solving for $\ell_z$ and inserting into our expression for $\gamma^2$ gives
	\begin{equation}
		\gamma_\text{max}^2 \approx \frac{8 \pi^{1/2}}{\nu_\text{crit} \mu_h^{1/2}} N \left(n_\text{3D}^\text{max} a_\text{eff}^3\right)^{1/2}.
	\end{equation}
	Therefore, with $10^5$ erbium or dysprosium atoms (for which $a_\text{dd} \approx 100a_0$, and setting $a_s \approx 0$), if we limit $n_\text{3D} \lesssim \SI{100}{\per\micro\metre\cubed}$, one could reach $\gamma_\text{max} \approx 40$ with $\ell_z \approx \SI{0.4}{\micro\metre}$ (equivalent to a vertical trapping frequency of $\approx \SI{400}{\hertz}$), resulting in $H \approx \SI{70}{\percent}$ for $p^*=8$ (cf.\ $H \approx \SI{10}{\percent}$ in a harmonic trap).
	
	In conclusion, we have explored the homogeneity of a dipolar gas, tuned close to its stability boundary, in a flattened, cylindrically symmetric power-law potential. We found that a large exponent in the power-law triggers density oscillations near the trap wall, which prevent the bulk of the trap achieving the density a perfectly homogeneous flattened system would have. An intermediate exponent is therefore more suitable, and we found its optimal value is determined by the trap wall steepness, which depends on both the aspect ratio and power-law exponent. These findings guide the way towards the experimental realisation of such a homogeneous dipolar gas for the study of, for example, droplet arrays, novel supersolid phases and critical phenomena.
	
	Data supporting this publication are openly available in Ref.~\cite{Juhasz:2022}.
	
	\begin{acknowledgments}
		We thank Tev\v{z} Lotri\v{c} for valuable discussions and Zoran Hadzibabic and Anna Marchant for comments on the manuscript. This work was supported by the UK EPSRC (grants no.\ EP/P009565/1 and EP/T019913/1). R.~P.~S.\ and P.~J.\ acknowledge support from the Royal Society, P.~J.\ acknowledges support from the Hungarian National Young Talents Scholarship and M.~K.\ from Trinity College, Cambridge.
	\end{acknowledgments}
	
	\appendix*
	\setcounter{equation}{0}

\section*{APPENDIX}
\label{app:numerics}
We numerically solve the GPE using both the preconditioned conjugate gradient method~\cite{Antoine:2017} and imaginary time propagation with the split-operator technique~\cite{Feit:1982} to cross-check our results. Our algorithm largely follows Ref.~\cite{Ronen:2006} with some differences laid out below, and is implemented in Python using several highly efficient and parallelised lower-level libraries for the most computationally expensive parts~\footnote{Matrix multiplications are performed using MKL via \texttt{numpy}, Fourier transforms are calculated via \texttt{mkl\_fft}. Piecewise array operations are executed using \texttt{numba}, a package which turns Python code into parallelised machine code.}. Given the trap has axial symmetry, the 3D problem can be reduced to a 2D one computationally. We sample the wave function on a grid and calculate the kinetic and dipolar interaction terms in the GPE using a Hankel transform along $r$ and a cosine transform along $z$, given the ground state is symmetric with respect to $z=0$. To avoid interaction between phantom copies of the cloud along $z$ due to the cosine transform, we employ a cutoff of the dipolar interaction in this direction (this is not a problem along $r$)~\cite{Ronen:2006}. The drawback of using a 2D grid is that it does not allow for instability due to angular excitations. To take these into account, we ensure that all angular excitations have a real positive energy using the Bogoliubov--de Gennes (BdG) formalism~\cite{Ronen:2006}. To find the largest interaction strength at which a stable condensate can be produced, we employ a binary search technique.

\subsection{Grid}
The grid needs to be large enough to comfortably contain the gas, whose size can be estimated using the Thomas--Fermi approximation. We calculate the Thomas--Fermi radii of a gas in our trap with an effective scattering length $a_\text{eff} = a_s+2a_\text{dd}$~\cite{Baillie:2015}, and find (in our dimensionless units)
\begin{equation}
	R_r = \left(\frac{3\nu_\text{crit}(\varepsilon_\text{dd})}{2}\right)^\frac{2}{3p} \gamma ,
	\quad
	R_z = \left(\frac{3\nu_\text{crit}(\varepsilon_\text{dd})}{2}\right)^\frac{1}{3} ,
\end{equation}
where $\nu_\text{crit}$ is tabulated in Ref.~\cite{Baillie:2015}.

Along $z$ we use a uniform grid with a grid size of 10, which is large enough to avoid interaction between phantom copies of the gas. Note that we use a constant grid size as $R_z$ depends only weakly on $\varepsilon_\text{dd}$.

The grid along $r$ is (slightly) non-uniform and is defined by $r_j=\alpha_j/\beta, j=0, \dotsc, N$ where $\alpha_j$ are the zeros of the first-order Bessel function $J_1(r)$~\cite{Kai-Ming:2009} and $\beta$ is chosen to give an overall grid size of $1.2 R_r$. The Hankel transform can be calculated on this grid with the same computational complexity as in Ref.~\cite{Ronen:2006}, but additionally it samples the centre of the trap.

This grid allows exact integration (for normalisation and to calculate the energy and the chemical potential) and interpolation (for expressing the wave function on different grids during the calculation of excitations). Similarly to Ref.~\cite{Ronen:2006}, for a function $f(r)$ sampled on this grid, it can be shown using a Dini series expansion that
\begin{equation}
	\int_0^\infty f(r) r \dd{r}=\frac{2}{\beta^2} \sum_{j=0}^N f(r_j) J_0^{-2}(\alpha_j).
\end{equation}
Exact integrals in $k$-space can be similarly calculated. Furthermore, like in Ref.~\cite{Ronen:2006}, using a Dini series expansion again an exact interpolation formula can be derived~\footnote{This formula is not well-defined for {$r=\alpha_j/\beta$}, but in that case {$r=r_j$} and the known {$f(r_j)$} can be directly used.}:
\begin{equation}
	f(r) = 2r\beta J_1(r\beta) \sum_{j=0}^N \frac{1}{r^2\beta^2-\alpha_j^2} J_0^{-1}(\alpha_j) f(r_j).
\end{equation}

The number of grid points are chosen to be $256 \times 65\ (r \times z)$ which ensures adequate sampling of the shortest relevant length scales (the oscillator length $\ell_z=1$ along $z$ and the roton wavelength $\lambda_\text{rot} \approx 2\pi$ along $r$). We checked that our results were insensitive to the exact number of grid points.

\subsection{Excitations}
Excitations can in general be written in the form $f(r,\theta,z) = f(r,z) e^{-im\theta}$~\cite{Ronen:2006}, where $f(r,z)$ has a definitive symmetry (even or odd) with respect to $z=0$ and $m$ is the phase winding number of the excitation. For a stable wave function, excitations with any $m$ must have a (real) positive energy and so excitations with a range of $m$ need to be calculated. The highest $m$ excitations that can lead to instability occur for high-$p$ traps, when the peak density is along a ring near the wall. In this case, the lowest-lying excitation can be thought of as a buckling along this ring (an angular roton~\cite{Ronen:2007}), such that $m_\text{crit} \approx 2\pi R_r / \lambda_\text{rot}$. In practice, we find testing above $1.25m_\text{crit}$ is not required and that always even excitations soften first, as the lowest-lying odd excitation is the Kohn mode, with exactly $\hbar \omega_z$ energy~\cite{Fetter:1998}.

We note that as the eigenvalues of the BdG equations are real for a stable condensate, to find the energy of the lowest-lying excitation it is sufficient to use the `SR' (smallest real part) mode of ARPACK~\cite{Ronen:2006}, without a preconditioner, which avoids calculating the inverse of the BdG matrix.

\subsection{Convergence}
To ensure our solution to the GPE has adequately converged, we:
\begin{enumerate}[label=(\alph*),nosep]
	\item required the smallest $m=0$ eigenvalue of the BdG equations to be (effectively) 0, as the presence of such a neutral mode confirms the ground state has been reached ~\cite{Ronen:2006};
	\item independently applied both the preconditioned conjugate gradient method~\cite{Antoine:2017} and imaginary time propagation with the split-operator technique~\cite{Feit:1982} and checked for consistency.
\end{enumerate}
However, as calculating BdG eigenstates is numerically expensive, we implemented less stringent but numerically much less expensive tests before the $m=0$ test takes place.

In the case of imaginary time propagation, convergence depends on both the time step size $\delta t$ and the criteria for halting the imaginary time propagation. For a given $\delta t$ we assume an exponential convergence (in imaginary time) of the wave function's energy, and require the energy difference from its infinite-time value to be below a certain threshold $E_\text{tol}$. By considering the change in $\ln(-\mathrm{d}E/\mathrm{d}t)$ in successive time steps, this energy difference can be calculated using the energy difference $\delta E_i$ between successive time steps, and the criteria amounts to requiring
\begin{equation}
	\frac{\delta E_i}{\ln(\frac{\delta E_{i}}{\delta E_{i-1}})} < E_\text{tol}.
\end{equation}
Making larger time steps $\delta t$ has the benefit of converging faster, but given the split-step method yields an error in the energy of $\order{\delta t^3}$, it comes at the expense of making larger errors. Therefore, after an iteration set with a certain $\delta t$ converged, we decrease our $\delta t$ by $\sqrt[3]{2}$ and continue with this procedure until the energy change between successive $\delta t$ iteration sets is also smaller than $E_\text{tol}$. We then do the $m=0$ BdG lowest eigenvalue test and lower $E_\text{tol}$ until it is passed.

For the preconditioned conjugate gradient method~\cite{Antoine:2017}, we follow the approach in Ref.~\cite{Antoine:2018}, with the choice of the combined (symmetric) preconditioner and the Polak--Ribi\`ere formula~\cite{Polak:1969} to enforce the conjugacy criterion. The convergence of this method is determined by a single threshold, by ensuring the energy change between subsequent ground state candidates is not more than $\delta E_\text{PCG} = 10^{-12}$. We found that this was sufficient to pass the $m=0$ test.
	
	\bibliography{References}
	
\end{document}